\documentstyle[twoside,fleqn,epsfig]{article}

\catcode`\@=11
\long\def\@makefntext#1{
\protect\noindent \hbox to 3.2pt {\hskip-.9pt
$^{{\eightrm\@thefnmark}}$\hfil}#1\hfill}               

\def\@makefnmark{\hbox to 0pt{$^{\@thefnmark}$\hss}}    

\def\ps@myheadings{\let\@mkboth\@gobbletwo
\def\@oddhead{\hbox{}
\rightmark\hfil\eightrm\thepage}
\def\@oddfoot{}\def\@evenhead{\eightrm\thepage\hfil
\leftmark\hbox{}}\def\@evenfoot{}
\def\sectionmark##1{}\def\subsectionmark##1{}}

\oddsidemargin=\evensidemargin
\newcounter{sectionc}\newcounter{subsectionc}\newcounter{subsubsectionc}
\renewcommand{\section}[1] {\vspace{12pt}\addtocounter{sectionc}{1}
\setcounter{subsectionc}{0}\setcounter{subsubsectionc}{0}\noindent
        {\tenbf\thesectionc. #1}\par\vspace{5pt}}
\renewcommand{\subsection}[1] {\vspace{12pt}\addtocounter{subsectionc}{1}
        \setcounter{subsubsectionc}{0}\noindent
        {\bf\thesectionc.\thesubsectionc. {\kern1pt \bfit #1}}\par\vspace{5pt}}
\renewcommand{\subsubsection}[1] {\vspace{12pt}\addtocounter{subsubsectionc}{1}
        \noindent{\tenrm\thesectionc.\thesubsectionc.\thesubsubsectionc.
        {\kern1pt \tenit #1}}\par\vspace{5pt}}
\newcommand{\nonumsection}[1] {\vspace{12pt}\noindent{\tenbf #1}
        \par\vspace{5pt}}

\newcounter{appendixc}
\newcounter{subappendixc}[appendixc]
\newcounter{subsubappendixc}[subappendixc]
\renewcommand{\thesubappendixc}{\Alph{appendixc}.\arabic{subappendixc}}
\renewcommand{\thesubsubappendixc}
        {\Alph{appendixc}.\arabic{subappendixc}.\arabic{subsubappendixc}}

\renewcommand{\appendix}[1] {\vspace{12pt}
        \refstepcounter{appendixc}
        \setcounter{figure}{0}
        \setcounter{table}{0}
        \setcounter{lemma}{0}
        \setcounter{theorem}{0}
        \setcounter{corollary}{0}
        \setcounter{definition}{0}
        \setcounter{equation}{0}
        \renewcommand{\thefigure}{\Alph{appendixc}.\arabic{figure}}
        \renewcommand{\thetable}{\Alph{appendixc}.\arabic{table}}
        \renewcommand{\theappendixc}{\Alph{appendixc}}
        \renewcommand{\thelemma}{\Alph{appendixc}.\arabic{lemma}}
        \renewcommand{\thetheorem}{\Alph{appendixc}.\arabic{theorem}}
        \renewcommand{\thedefinition}{\Alph{appendixc}.\arabic{definition}}
        \renewcommand{\thecorollary}{\Alph{appendixc}.\arabic{corollary}}
        \renewcommand{\theequation}{\Alph{appendixc}.\arabic{equation}}
        \noindent{\tenbf Appendix \theappendixc #1}\par\vspace{5pt}}
\newcommand{\subappendix}[1] {\vspace{12pt}
        \refstepcounter{subappendixc}
        \noindent{\bf Appendix \thesubappendixc. {\kern1pt \bfit #1}}
        \par\vspace{5pt}}
\newcommand{\subsubappendix}[1] {\vspace{12pt}
        \refstepcounter{subsubappendixc}
        \noindent{\rm Appendix \thesubsubappendixc. {\kern1pt \tenit #1}}
        \par\vspace{5pt}}

\newcommand{\textlineskip}{\baselineskip=13pt}
\newcommand{\smalllineskip}{\baselineskip=10pt}

\def\eightcirc{
\begin{picture}(0,0)
\put(4.4,1.8){\circle{6.5}}
\end{picture}}
\def\eightcopyright{\eightcirc\kern2.7pt\hbox{\eightrm c}}

\newcommand{\copyrightheading}[1]
        {\vspace*{-2.5cm}\smalllineskip{\flushleft
        {\footnotesize International Journal of Modern Physics D, #1}\\
        {\footnotesize $\eightcopyright$\, World Scientific Publishing
         Company}\\
         }}

\newcommand{\publisher}[2]{{\begin{center}\footnotesize\smalllineskip
        Received #1\\
        Revised #2
        \end{center}
        }}

\def\abstracts#1#2#3{{
        \centering{\begin{minipage}{4.5in}\baselineskip=10pt\footnotesize
        \parindent=0pt #1\par
        \parindent=15pt #2\par
        \parindent=15pt #3
        \end{minipage}}\par}}

\def\keywords#1{{
        \centering{\begin{minipage}{4.5in}\baselineskip=10pt\footnotesize
        {\footnotesize\it Keywords}\/: #1
         \end{minipage}}\par}}

\renewenvironment{thebibliography}[1]
        {\frenchspacing
         \ninerm\baselineskip=11pt
         \begin{list}{\arabic{enumi}.}
        {\usecounter{enumi}\setlength{\parsep}{0pt}
         \setlength{\leftmargin 12.7pt}{\rightmargin 0pt} 
         \setlength{\itemsep}{0pt} \settowidth
        {\labelwidth}{#1.}\sloppy}}{\end{list}}

\newcounter{itemlistc}
\newcounter{romanlistc}
\newcounter{alphlistc}
\newcounter{arabiclistc}

\newcommand{\fcaption}[1]{
        \refstepcounter{figure}
        \setbox\@tempboxa = \hbox{\footnotesize Fig.~\thefigure. #1}
        \ifdim \wd\@tempboxa > 5in
           {\begin{center}
        \parbox{5in}{\footnotesize\smalllineskip Fig.~\thefigure. #1}
            \end{center}}
        \else
             {\begin{center}
             {\footnotesize Fig.~\thefigure. #1}
              \end{center}}
        \fi}

\newcommand{\tcaption}[1]{
        \refstepcounter{table}
        \setbox\@tempboxa = \hbox{\footnotesize Table~\thetable. #1}
        \ifdim \wd\@tempboxa > 5in
           {\begin{center}
        \parbox{5in}{\footnotesize\smalllineskip Table~\thetable. #1}
            \end{center}}
        \else
             {\begin{center}
             {\footnotesize Table~\thetable. #1}
              \end{center}}
        \fi}

\def\@citex[#1]#2{\if@filesw\immediate\write\@auxout
        {\string\citation{#2}}\fi
\def\@citea{}\@cite{\@for\@citeb:=#2\do
        {\@citea\def\@citea{,}\@ifundefined
        {b@\@citeb}{{\bf ?}\@warning
        {Citation `\@citeb' on page \thepage \space undefined}}
        {\csname b@\@citeb\endcsname}}}{#1}}

\newif\if@cghi
\def\cite{\@cghitrue\@ifnextchar [{\@tempswatrue
        \@citex}{\@tempswafalse\@citex[]}}
\def\citelow{\@cghifalse\@ifnextchar [{\@tempswatrue
        \@citex}{\@tempswafalse\@citex[]}}
\def\@cite#1#2{{$\null^{#1}$\if@tempswa\typeout
        {IJCGA warning: optional citation argument
        ignored: `#2'} \fi}}

\def\pmb#1{\setbox0=\hbox{#1}
        \kern-.025em\copy0\kern-\wd0
        \kern.05em\copy0\kern-\wd0
        \kern-.025em\raise.0433em\box0}

\def\fnt#1#2{\footnotetext{\kern-.3em
        {$^{\mbox{\scriptsize #1}}$}{#2}}}

\def\fpage#1{\begingroup
\voffset=.3in
\thispagestyle{empty}\begin{table}[b]\centerline{\footnotesize #1}
        \end{table}\endgroup}

\def\runninghead#1#2{\pagestyle{myheadings}
\markboth{{\protect\footnotesize\it{\quad #1}}\hfill}
{\hfill{\protect\footnotesize\it{#2\quad}}}}
\font\tenrm=cmr10
\font\tenit=cmti10
\font\tenbf=cmbx10
\font\bfit=cmbxti10 at 10pt
\font\ninerm=cmr9

\font\eightrm=cmr8

\def\qed{\hbox{${\vcenter{\vbox{                        
   \hrule height 0.4pt\hbox{\vrule width 0.4pt height 6pt
   \kern5pt\vrule width 0.4pt}\hrule height 0.4pt}}}$}}

\begin{document}

\runninghead{Black Holes of a Minimal Size
$\ldots$} {$\ldots$ in String Gravity}

\normalsize\textlineskip
\thispagestyle{empty}
\setcounter{page}{1}

\copyrightheading{}             

\vspace*{0.88truein}

\fpage{1}

\centerline{\bf BLACK  HOLES OF A MINIMAL  SIZE}
\vspace*{0.035truein}
\centerline{\bf IN STRING GRAVITY}
\vspace*{0.37truein}
\centerline{\footnotesize
S.O. Alexeyev${}^1$\footnote{Electronic address: alexeyev@grg2.phys.msu.su},
M.V. Sazhin${}^{1,3}$\footnote{Electronic address: sazhin@sai.msu.su},
M.V. Pomazanov${}^2$\footnote{Electronic address: michael@math356.phys.msu.su}}
\vspace*{0.015truein}

\centerline{\footnotesize\it
${}^1$ Sternberg Astronomical Institute,
Moscow State University,}

\centerline{\footnotesize\it
Universitetskii Prospect, 13, Moscow 119899, RUSSIA}

\centerline{\footnotesize\it
${}^2$ Department of Mathematics, Physics Faculty,}

\centerline{\footnotesize\it
Moscow State University, Moscow 119899, RUSSIA}

\centerline{\footnotesize\it
${}^{3}$ Isaac Newton Institute for Mathematical Sciences,
         University of Cambridge, UK}
\vspace*{0.225truein}
\publisher{(received date)}{(revised date)}

\baselineskip=10pt

\vspace*{0.21truein}
\abstracts{
     A lower limit for a neutral black hole size is  obtained in
the frames of  the string gravity  model with the  second  order
curvature correction. It  is shown that this effect remains when
the third order  curvature correction is also taken into account
and argued that  such restriction does exist in all perturbative
orders of curvature expansions.}{}{}

\vspace*{10pt}
\keywords{black hole, string theory, higher order curvature corrections}

\vspace*{1pt}\textlineskip

\noindent
     There  are  a  lot  of unsolved (and  even  non-understood)
problems in the modern theoretical physics now. One  of the most
intriguing of them is a question on the endpoint of a black hole
evaporation \cite{h1}.  This  problem  is  widely  discussed now
\cite{th2}  because   a  complete  evaporation  of  black  holes
(without  any  remnants) can violate the Quantum Coherence.  The
indirect indications  to  a  possible  existence  of the Quantum
Coherence violation were found early \cite{saz1}. This puzzle is
real  only  for  black  holes  with  initial  masses  less  than
$10^{15}$ grams  \cite{c3}.  Hitherto  such  black  holes  could
either completely evaporate or  they  could survive as some real
objects. The subject of the study is to find a  model describing
their form of  an existence. General Relativity does not suggest
any   variants   of  the  solution  of  this  puzzle.   Perhaps,
non-perturbative  M-theory,  presently  developed,  can  clarify
something but it is only at the beginning of its way. Therefore,
in order to make a little step in this direction, it is possible
(working in  quasi-classical  approach)  to  use the non-minimal
gravity model which represents the effective low energy limit of
some great unification theory.

     {\it In the perturbational approach} the  string theory (as
a  part  of  M-theory)  predicts  the  Einstein equations to  be
modified by  higher  order  curvature  corrections  in the range
where the curvature of space-time has  the near-Planckian values
\cite{gws}. In this approach of the string theory one has to use
the so-called effective  low energy action with the higher order
curvature  corrections  in  order  to extend the  boundaries  of
applicability of  the classical General Relativity. The addition
of the next curvature  term allows one to make the next  step in
this  direction.  But all  the  resulting  conclusions  must  be
treated as preliminary directions on  conclusions  in  the  near
Planckian region, later they have to be checked  by pure quantum
calculations.  At  the  present  time the form of  higher  order
curvature corrections  in  the  string  effective  action is not
completely investigated \cite{e13}.  We  do not know the general
structure of the expansion and, hence, the direct  summing up is
impossible.  But  as  we  deal  with  the  expansion,  the  most
important correction is  the second order curvature one which is
the  product  of  the  Gauss-Bonnet  and   dilatonic  terms.  As
Gauss-Bonnet term  is  a  total  divergence  in four dimensional
space-time,   its   combination   with    a    dilatonic   field
($e^{-2\phi}L_2$, see  below)  provides  only  the  second order
differential equations relatively metric. Thus, the generic form
of the action  for all kinds  of strings (for  simplicity,  only
bosonian part is taken into account) has the form
\begin{eqnarray}\label{e1.1} S & = &
\frac{1}{16\pi} \int d^4 x \sqrt{-g}
 \biggl[
m^2_{Pl} \biggl( -R + 2 \partial_{\mu} \phi
\partial^\mu \phi \biggr)
+ \lambda_2 e^{-2\phi} L_2 \biggr],
\end{eqnarray}
where
\begin{eqnarray}
L_2 & = & R_{ijkl}R^{ijkl} - 4 R_{ij}R^{ij} + R^2 . \nonumber
\end{eqnarray}
     Here    $\phi$     is     the     dilatonic    field    and
$\lambda_2=\lambda_2(\alpha')$ is the  string coupling constant.
Its value depends upon the type of the string theory.

     The investigations in the frames  of  the  discussed  model
were performed formerly \cite{e7,e9}. One of  its most important
results is the determination  of  the restriction to the minimal
black hole mass.  This effect was  found both in  the  numerical
calculations and in the analytical ones and was independent from
the metric parametrization. In Plank unit values it has the form
\begin{equation}\label{e9}
r_h^{inf} = \sqrt{\lambda} \ 2 \ {}^4\sqrt{6},
\end{equation}
     where $r_h$  is the {\it  lower limit} value of the horizon
radius  (for  more  details  see  \cite{e9}).  This  restriction
appears in  the second order  curvature gravity and is absent in
the minimal Einstein-Schwarzschild gravity.

     In order  to  understand  whether  the  formula  (\ref{e9})
represents the fundamental restriction resulting from the string
theory or  it is only the  effect of the  Gauss-Bonnet curvature
correction, it is  necessary to investigate the situation in the
following perturbation orders. The general form of the action is
\cite{e13}
\begin{eqnarray}\label{act2}
S & = & \frac{1}{16\pi}
\int d^4 x \sqrt{-g} \biggl[ - m^2_{Pl} R
+ 2 \partial_{\mu} \phi
\partial^{\mu} \phi \nonumber \\
& + & \lambda_2 e^{-2 \phi} L_2
+ \lambda_3 e^{-4 \phi} L_3
+ \lambda_4 e^{-6 \phi} L_4 + \ldots \biggr].
\end{eqnarray}
     Here $L_2$  denotes  the  second order curvature correction
(Gauss-Bonnet term),  $L_3$  denotes  the  third order curvature
correction,
\begin{eqnarray}
L_3 & = & R^{\mu\nu}_{\alpha\beta} R^{\alpha\beta}_{\lambda\rho}
R^{\lambda\rho}_{\mu\nu} + 2 \Omega_3, \nonumber
\end{eqnarray}
where
\begin{eqnarray}
\Omega_3 & = &
R^{\mu\nu}_{\alpha\beta} R^{\alpha\beta}_{\lambda\rho}
R^{\lambda\rho}_{\mu\nu}
- 2 R^{\mu\nu}_{\alpha\beta} R_\nu^{\lambda\beta\rho}
R^\alpha_{\rho\mu\lambda}
+ \frac{3}{4} R R^2_{\mu\nu\alpha\beta} \nonumber \\
& + & 6 R^{\mu\nu\alpha\beta} R_{\alpha\mu} R_{\beta\nu}
+ 4 R^{\mu\nu} R_{\nu\alpha} R^\alpha_\mu - 6 R R^2_{\alpha\beta}
+ \frac{1}{4} R^3 , \nonumber
\end{eqnarray}
     and $L_4$  denotes  the  fourth order curvature correction.
The rest  members  of  formula  (\ref{act2})  are $\lambda_3=c_3
\lambda_2^2$,   $\lambda_4=c_4    \lambda_2^3$,   $\ldots$   The
coefficients $c_i$ depend upon the type  of  the  string  theory
(complete set of their values can be found in \cite{e13}).

     Similar to our previous work \cite{e9},  we  look  for  the
static,  spherically symmetric,  asymptotically  flat  solutions
providing   a   regular    (``quasi-Schwarzschild'')    horizon.
Therefore,  the  most convenient  choice  of  metric  (which  is
usually called as the ``curvature gauge'') is
\begin{eqnarray}\label{e1.2}
ds^2 = \Delta dt^2 - \frac{\sigma^2 }{\Delta } dr^2 - r^2
(d \theta^2 + \sin^2 \theta d \varphi^2),
\end{eqnarray}
     where $\Delta=\Delta(r)$, $\sigma=\sigma(r)$. The curvature
gauge and the  Einstein  frame are  used  for a more  convenient
comparison with the previous results.

     Field equations  obtained from the action (\ref{act2}) have
the form
\begin{eqnarray}\label{eq1}
\biggl[ A_1 + \lambda_2 A_2 \biggr] \left(
\begin{array}{c}
\Delta'' \\ \sigma' \\ \phi''
\end{array}
\right)
= B_1 + \sum\limits_{i=2}^{\infty} \lambda_i B_i ,
\end{eqnarray}
where $(3\times 3)$ matrix $A_1=A_1(\Delta,\Delta',\sigma,\phi',r)$
and the column-vector \\
$B_1=B_1(\Delta,\Delta',\sigma,\phi',r)$
come from the variation of the Einstein part of the action
(\ref{act2}) \cite{e9}.
$(3\times 3)$ matrix $A_2=A_2(\Delta,\Delta',\sigma,\phi,\phi',r)$
and the column-vector $B_2=B_2(\Delta,\Delta',\sigma,\phi,\phi',r)$
are the result of the variation
of the Gauss-Bonnet part of the action
(\ref{act2}) \cite{e9}. Column-vectors $B_3$, ($B_4$, $\ldots$),
$B_i=B_i(\Delta \ldots\Delta'''',
\sigma\ldots\sigma''',\phi\ldots\phi'',r)$
are the consequence  of the variation of the third (fourth,
$\ldots$) order  curvature  correction.  Unfortunately, due to a
huge size  of $B_i$ expressions,  we have no opportunity to show
them in this  letter.  Here  it is necessary to  note  that  the
curvature  corrections have the  general  form  $R  * R  *  R  *
\ldots$. So, they do not  produce  the  highest derivatives with
increasing  order  in  the  field  equations  relatively  $L_3$.
Therefore,   the   contribution   of  $B_3$  ($B_4$,   $\ldots$)
represents the singular perturbation for the $(A_1,A_2,B_1,B_2)$
part of the Eq. (\ref{eq1}).

     Let us consider the  question  about the influence of $B_3$
part   to   the  solution  of  the  equations  consisting   from
$(A_1,A_2,B_1,B_2)$   terms.  The   main   mathematical   reason
providing the existence  of the restriction to the minimal black
hole size is the intersection of the solution  with the singular
surface defined  by  $\mbox{det}(A_1  +  \lambda_2 A_2)=0$ which
contains  a  caustic.  Since  in  the   infinity  the  curvature
corrections vanish more quickly than the Einstein part, since in
the infinity the asymptotic solution is completely determined by
the Arnowitt-Deser-Misner mass $M$ and the  dilatonic charge $D$
(here we define $\phi_\infty=0$ for simplicity). In other words,
the  initial  data  for  the  numerical   calculations  for  all
functions $\Delta$, $\sigma$ and $\phi$ are
\begin{eqnarray}\label{eq2}
\Delta & = & 1 - \frac{2M}{r}
+ O\biggl(\frac{1}{r}\biggr), \nonumber \\
\sigma & = & 1 - \frac{1}{2} \frac{D^2}{r^2}
+ O\biggl(\frac{1}{r^2}\biggr), \\
\phi   & = & \frac{D}{r}
+ O\biggl(\frac{1}{r}\biggr). \nonumber
\end{eqnarray}
     The last asymptotics suppose  that  in a case of $\lambda_3
\rightarrow  0$,  the solution of Eqs. (\ref{eq1}) converges  to
the solution of  Eqs. (\ref{eq2}). This occurs when the solution
of Eqs. (\ref{eq1}) perturbed by the term $B_3$ with the highest
derivatives  is  regular (boundary layers are absent). In  other
words,  in  the  case  $\lambda_3 \rightarrow 0$  this  solution
converges   to   the   regular   part   of   the   main   branch
$[A_1,A_2,B_1,B_2]$.  Consequently,   one   has   to  check  the
stability of  the $[A_1,A_2,B_1,B_2]$ solution relatively to the
regular perturbation of the singular contribution $B_3$.

     Eqs. (\ref{eq1})  with  the  contribution $B_3$ were solved
numerically. The solution was obtained by  the iteration method.
At every step system  (\ref{eq1})  was solved by the Runge-Kutta
method  (see  \cite{e9}  for  more  details   of  our  numerical
integration strategy),  the highest derivatives being taken from
the previous  step. Figure 1  shows the dependence of the metric
function  $\Delta$  against  the  radial coordinate $r$  at  the
different values  of the event horizon  $r_h$ (in the  case when
$\lambda_2=\lambda_3=1$).  Similar singularity  case  was  found
recently by K.Maeda et  al.  \cite{e8}. The curve (a) represents
the case when $r_h$ is rather  large and is equal to 20.0 Planck
unit values (P.u.v.). The curve (b) shows $\Delta(r)$ with $r_h$
being rather close to ${r_h}_{min}$. The  curve  (c)  shows  the
case when  $2M \ll {r_h}_{min}$ and  any horizon is  absent. One
should make a note that the value of ${r_h}_{min}$ has  the same
order as in the Gauss-Bonnet case.

     Here we  would like to  add that the asymptotic behavior of
the   discussed    solution    near    the   horizon   is   also
``quasi-Schwarzschild'' one  and, as we checked numerically, the
restriction to  a minimal black hole  size can be  obtained from
these asymptotic expansions.

     From  our  analysis  we  can conclude that when  the  third
curvature  correction   $L_3$   is   taken   into  account,  the
restriction  to  the minimal black hole size  does  retain.  Its
numerical value slightly  differs from the second order one, but
the  effect  retains  principally.  In so far as  the  curvature
corrections $L_i$,  $i=4,\ldots$  do  not  produce  the  highest
derivatives with the  increasing  order relatively $L_3$ one, so
far as {\it a  particularity  existing in the third perturbation
order exists in all perturbative orders. Thus, it is possible to
get a conclusion that  the  discussed restriction to the minimal
black  hole  mass  is  a fundamental restriction of  the  string
theory because it  takes place in all perturbative orders.} Note
that after taking  into  account the whole curvature expansions,
one  has  a  right to work with the masses of Plank order in the
quasi-classical level.

     {\bf Conclusions}

     We have analyzed the four dimensional  black hole solutions
appearing in the  low  energy limit in all  kinds  of the string
theory. When the higher order curvature  corrections are allowed
for the analysis,  the restriction to  the minimal size  of  the
black hole appears. For example,  it  is  approximately equal to
$0.4\ m_{Pl}$ in the second order  for  the  heterotic  strings.
Such  restriction  does exist  in  all  perturbation  orders  of
curvature expansions. Here it is necessary to point out that the
same result was obtained  during  the quantization of black hole
with selfgraviting dust shell \cite{berezin}.

     And finally, speculating on this phenomenon (if this object
is  stable),  the  minimal (``quasi-Schwarzschild'') black  hole
being the endpoint  of the Hawking evaporation can represent the
relic remnant of black holes formed during the initial stages of
our Universe formation. This is the very interesting problem and
it requires the future investigations.

     {\bf Acknowledgments}

     The authors are  indebted  to Profs. V.Rubakov, K.Maeda and
Dr. S.Mignemi  for  valuable  discussions.  M.Sazhin is indebted
Isaac Newton  Institute for Mathematical Sciences, University of
Cambridge, for  hospitality.  S.Alexeyev  would  like  to  thank
Universita di  Cagliari,  Dipartimento  di  Matematica and INFN,
Sezione di Cagliary for hospitality.

\nonumsection{References}

\pagebreak

\begin{figure}
\hspace*{-2cm}
\vspace*{1cm}
\epsfig{figure=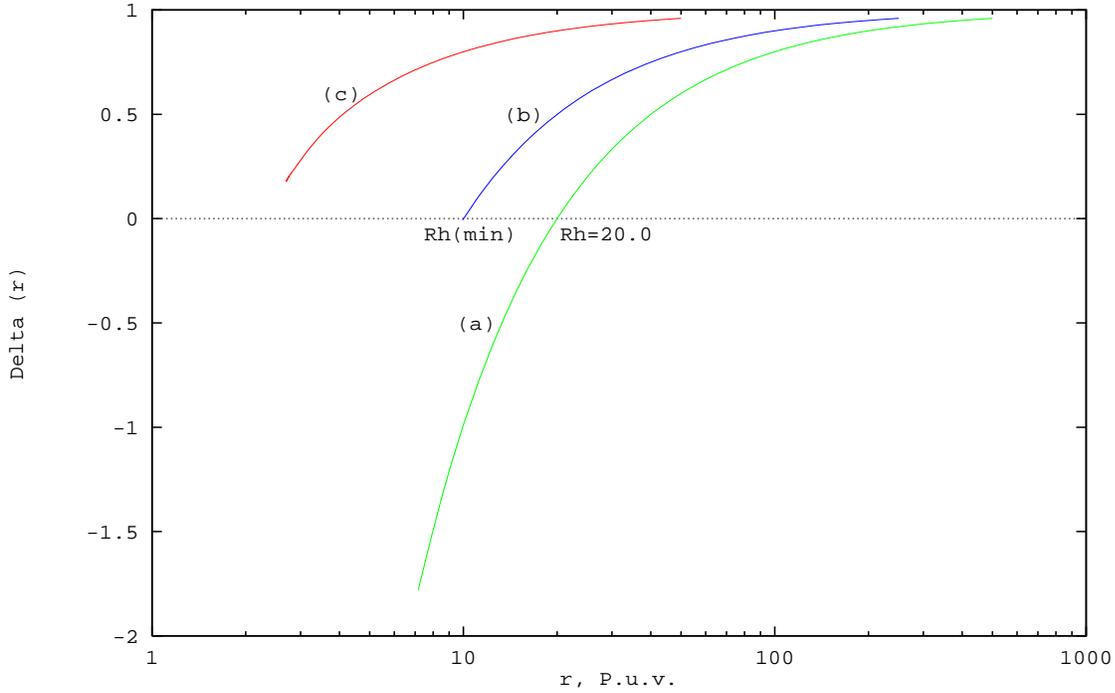, width=15cm, height=10cm}
\vspace*{5mm}
\caption{
     The dependence of the metric function  $\Delta$ against the
radial  coordinate  $r$  at  the different values of  the  event
horizon $r_h$  (in  the  case when $\lambda_2=\lambda_3=1$). The
curve (a) represents the case when $r_h$ is rather large  and is
equal to  20.0 Planck unit  values (P.u.v.). The curve (b) shows
$\Delta(r)$ with $r_h$ being rather close  to ${r_h}_{min}$. The
curve  (c) shows  the  case when $2M  \ll  {r_h}_{min}$ and  any
horizon is absent.
}
\end{figure}

\end{document}